\documentclass[aps,pre,reprint,showpacs,preprintnumbers,amsmath,amssymb]{revtex4-1}

\usepackage{amsfonts,dsfont,bm}
\usepackage{graphicx}
\usepackage[utf8x]{inputenc}
\usepackage{natbib}
\usepackage[colorlinks=true,citecolor=blue,linkcolor=blue]{hyperref}

\newcommand{\run}{\ensuremath{\mathbf{r_1}}}
\newcommand{\rde}{\ensuremath{\mathbf{r_2}}}
\newcommand{\Rv}{\ensuremath{\mathbf{R}}}
\newcommand{\rv}{\ensuremath{\mathbf{r}}}
\newcommand{\qv}{\ensuremath{\mathbf{q}}}
\newcommand{\kv}{\ensuremath{\mathbf{k}}}
\newcommand{\diag}[1]{\begin{minipage}[h!]{1.351cm}\includegraphics[width=1.35cm]{#1}\end{minipage}}

\begin{document}

\title{Fluctuations of the Casimir-like force between two membrane inclusions}
\author{Anne-Florence Bitbol}
\author{Paul G. Dommersnes} 
\author{Jean-Baptiste Fournier}
\affiliation{Laboratoire Mati\`ere et Syst\`emes Complexes (MSC),
Universit\'e Paris Diderot, Paris~7 and UMR CNRS 7057, 10 rue Alice Domon
et L\'eonie Duquet, F-75205 Paris Cedex 13, France}
\date{\today}

\begin{abstract}
Although Casimir forces are inseparable from their fluctuations, little is known about these fluctuations in soft matter systems. We use the membrane stress tensor to study the fluctuations of the membrane-mediated Casimir-like force. This method enables us to recover the Casimir force between two inclusions and to calculate its variance. We show that the Casimir force is dominated by its fluctuations. Furthermore, when the distance $d$ between the inclusions is decreased from infinity, the variance of the Casimir force decreases as $-1/d^2$. This distance dependence shares a common physical origin with the Casimir force itself.
\end{abstract}

\maketitle

In 1948, Casimir predicted that two uncharged metallic plates placed in vacuum should attract each other \cite{Casimir48}. 
Indeed, the boundary conditions imposed by the plates on the electromagnetic field constrain its quantum fluctuation modes in such a way that the zero-point energy of the system depends on the distance between the plates. 
Such long-range fluctuation-induced forces arise between any objects that impose boundary conditions on a fluctuating field with long-range correlations \cite{Kardar99}. 
Casimir-like effects driven by thermal fluctuations of material media were first discussed by Fisher and de Gennes in the context of critical mixtures \cite{Fisher78, Krech94,Gambassi09}. They also appear in soft matter systems such as liquid crystals \cite{Ajdari91, Ziherl00}, fluid membranes \cite{Goulian93etc, Park96, Dommersnes99, Dommersnes99b, Dean06} and fluid interfaces \cite{Noruzifar09}, as well as in superfluids \cite{Li92,*Maciolek07}. The first direct measurement of thermal Casimir-like forces was achieved very recently in a critical binary mixture by Hertlein et al. \cite{Hertlein08}. Experimental evidence for fluctuation-induced forces between lipid domains in vesicles was also provided very recently \cite{Semrau09}.

The Casimir force actually coincides with the average of the stress exerted by the fluctuating field \cite{Brown69}. Although Casimir forces are by essence inseparable from their fluctuations, the latter have been scarcely studied. In 1991, Barton first characterized the fluctuations of the quantum Casimir force by calculating the variance of the stress tensor \cite{Barton91}. There have been few studies since then \cite{Jaekel92,*Robaschik95,*Messina07}. The fluctuations of Casimir forces are, however, of fundamental importance. Indeed, Casimir force measurements are always performed by probing a fluctuating quantity, either the force itself \cite{Lamoreaux97, Mohideen98} or the position of one of the interacting objects \cite{Hertlein08}. In addition, the distance dependence of the fluctuations of Casimir forces is intriguing as it shares a common origin with the Casimir effect: the suppression of fluctuating degrees of freedom. 

The study of the fluctuations of Casimir-like forces was initiated by Bartolo et al., who considered the case of parallel plates imposing Dirichlet boundary conditions on a fluctuating scalar field \cite{Bartolo02}. Although the case studied in Ref. \cite{Bartolo02} is quite generic, including, e.g., classical spin systems \cite{Dantchev04}, it does not encompass all the soft matter systems that can give rise to a Casimir-like effect. Membrane inclusions, for instance, which are small objects interacting with the membrane through a complicated stress tensor \cite{Capovilla02, Fournier07}, are left out.

In this paper, we calculate the fluctuations of the membrane-mediated Casimir-like force acting between two inclusions, e.g., proteins, that locally constrain the curvature of the membrane. First, we show that the Casimir-like force, which is usually derived from the membrane free energy \cite{Goulian93etc, Park96, Dommersnes99, Dommersnes99b}, can be obtained by integrating the average membrane stress tensor. Then we calculate the variance of this force and we study its distance dependence. We also discuss the effect on the variance of the interplay between the Casimir-like force and the curvature-dependent force.

\textit{Model.}---Biological membranes are fluid lipid bilayers that are well described on a large scale by an elastic curvature energy known as the Helfrich Hamiltonian \cite{Helfrich73}:
\begin{equation}
H=\int\!\mathrm{d}\mathbf{r}\,\,\frac{\kappa}{2} \left[\nabla^2 h(\rv)\right]^2,
\label{H}
\end{equation}
where $h(\mathbf{r})$, with $\mathbf{r}=(x,y)$, denotes the height of the membrane with respect to a reference plane, and $\kappa$ is the membrane bending rigidity. This form holds for symmetric membranes, constituted of two identical monolayers, undergoing small deformations with respect to the flat shape and for distances smaller than $\sqrt{\kappa/\sigma}$, where $\sigma$ is the membrane tension.
The components of the stress tensor $\mathbf{\Sigma}$ associated with the effective Hamiltonian $H$ are given by \cite{Fournier07}:
\begin{eqnarray}
\Sigma_{xx}&=&\frac{\kappa}{2} \left[\left(\partial_y^2 h\right)^2-\left(\partial_x^2 h\right)^2\right]+\kappa \left(\partial_x h\right)\partial_x \nabla^2 h\,, \label{Sigmaxx}\\
\Sigma_{xy}&=&\kappa \left[\left(\partial_x h\right)\partial_y \nabla^2 h-\left(\partial_x \partial_y h\right)\nabla^2 h\right]. \label{Sigmaxy}
\end{eqnarray}

We consider inclusions that locally constrain the curvature of the membrane. This corresponds to the most generic case for inclusions not subject to external forces or torques. Particles such as proteins with Bin/amphiphysin/Rvs (BAR) domains \cite{Blood06} and viral capsids \cite{Gottwein03} are well-known examples of inclusions producing local membrane curvature \cite{McMahon05, Zimmerberg06, Reynwar07}. 

\begin{figure}[t b]
  \begin{center}
    \includegraphics[width=.60\columnwidth]{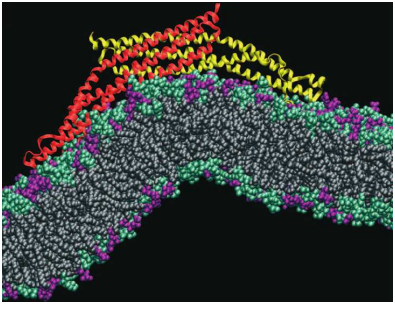}
    \caption{(Color online). Atomistic molecular dynamics simulation of a BAR domain inducing a strong local curvature in a lipid membrane. Reproduced with permission from Ref.~\cite{Blood06}. Copyright (2006) National Academy of Sciences, USA.}
  \label{BAR}
  \end{center}      
\end{figure}

We model the inclusions as point-like objects. This is justified since the typical radius of membrane proteins is comparable to the membrane thickness (see Fig.~\ref{BAR}), which vanishes in our coarse-grained description. This simplification makes the calculation of the force variance tractable. A disadvantage of this model, however, is that the size of the inclusions and the ultraviolet cutoff are not independent from one another. 

\textit{Average force.}---We first consider the case of two inclusions imposing a vanishing curvature, since it gives the Casimir-like force without the curvature-dependent contribution that would arise from an average deformation of the membrane \cite{Goulian93etc, Dommersnes99, Dommersnes99b}.
In order to calculate the average of the stress tensor, we need the correlation function of the membrane height, and therefore the partition function. The latter is obtained by integrating over all the configurations satisfying the constraints $\partial^2_x h=\partial_x \partial_y h=\partial^2_y h=0$ at the positions $\run$ and $\rde$ of the two inclusions:
\begin{equation}
Z[u]=\int\! \mathcal{D}h\,\,\prod_{i=1}^6\delta\left(\mathds{D}_i h (\mathbf{R}_i)\right)\,e^{-\beta H +\int\! d\rv \,\, h(\rv)u(\rv)}.
\end{equation}
Here, $\beta=1/k_\mathrm{B}T$ and $u(\mathbf{r})$ is an external field conjugate to $h(\mathbf{r})$; we have used the vectors $\mathds{D}=(\partial^2_x$, $\partial_x \partial_y$, $\partial^2_y$, $\partial^2_x$, $\partial_x \partial_y$, $\partial^2_y)$ and $\mathbf{R}=(\run$, $\run$, $\run$, $\rde$, $\rde$, $\rde)$ to express the constraints. Writing the Dirac distributions as Fourier transforms and integrating first over $h(\mathbf{r})$ and then over the Fourier variables yields \cite{Dommersnes99, Dommersnes99b}:
\begin{equation}
Z[u]=Z[0]\exp\left[ \frac{1}{2}\!\int\!d\mathbf{r}\,d\mathbf{r}'\, u(\rv)\,C(\rv,\rv')\,u(\rv') \right],
\label{Z}
\end{equation}
where $C(\rv,\rv')=\langle h(\rv)\,h(\rv')\rangle=G(\rv-\rv')+G'(\rv,\rv')$ is the correlation function of the membrane height. $ G(\rv)=\int_\mathbf{q} e^{i\qv\cdot\rv}/q^4$ is the Green function of the operator associated with the Helfrich Hamiltonian (\ref{H}). Here and in the following, we use the shorthand $\int_\mathbf{q}\equiv(k_\mathrm{B}T/\kappa)\int_{1/\xi\le|\mathbf{q}|\le\Lambda}d\mathbf{q}/(2\pi)^2$, where $\Lambda$ is the ultraviolet cutoff, comparable with the inverse membrane thickness, and $\xi$ stands for the membrane size in the absence of tension, or the fluctuation correlation length $\xi\approx\sqrt{\kappa/\sigma}$ otherwise. The second term in the correlation function, which comes from the presence of the inclusions, is found to be
\begin{equation}
G'(\rv,\rv')=-\!\sum_{i,j=1}^6 \left[\mathds{D}_i G(\rv-\Rv_i)\right]M^{-1}_{ij}\,\mathds{D}_j G(\rv'-\Rv_j),
\end{equation}
where $M_{ij}=\mathds{D}_i \mathds{D}_j G(\Rv_i-\Rv_j)$.

\begin{figure}[tb]
  \begin{center}
    \includegraphics[width=.75\columnwidth]{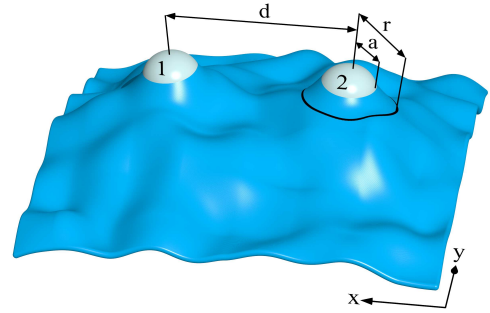}
    \caption{(Color online). Two curvature-inducing inclusions of radius $a$ separated by a distance $d$ in a fluctuating membrane. The black line is the contour used for the calculation of the Casimir force by integration of the stress tensor.}
  \label{schema_casimir}
  \end{center}      
\end{figure}

Let $\bm{f}$ be the force exerted by the rest of the membrane on a circular membrane patch of radius $r$ centered on the inclusion number 2 (see Fig.~\ref{schema_casimir}). Its component along the $x$ axis joining the two inclusions is 
\begin{equation}
f_x=r\int_{0}^{2\pi}\! d\theta\, \Big[\Sigma_{xx}(\rv) \cos\theta+\Sigma_{xy}(\rv)\sin\theta\Big],
\label{force}
\end{equation}
with $\rv=r\,(\cos\theta,  \sin\theta)$. Let us calculate the average of $f_x$. Thanks to Eq.~(\ref{Sigmaxx}--\ref{Sigmaxy}), $\langle\Sigma_{xx}\rangle$ and $\langle\Sigma_{xy}\rangle$ can be expressed as linear combinations of the derivatives of the correlation function $C$. For instance, the first term of $\Sigma_{xx}$ gives $\langle(\partial_y^2 h (\rv))^2\rangle=\partial_y^2\partial_{y'}^2 C(\rv,\rv')|_{\rv'=\rv}$. The divergence of the average stress tensor vanishes everywhere except on the inclusions, which means that $\langle f_x\rangle$ is independent of the contour chosen as long as it surrounds one inclusion. Indeed, for all $r$ smaller than the distance $d$ between the inclusions, we obtain the Casimir-like force
\begin{equation}
\left\langle f_x\right\rangle= 24\,k_\mathrm{B}T \,\frac{(2/\Lambda)^4}{d^5}+\mathcal{O}\left((\Lambda d)^{-9}\right),
\label{FCas}
\end{equation}
which was first derived from the membrane free energy \cite{Park96, Dommersnes99, Dommersnes99b}. 
This result coincides with the Casimir-like force between two rigid disks of radius $a$ \cite{Goulian93etc} for $\Lambda=2/a$ \cite{Park96, Dommersnes99, Dommersnes99b}.
Thus, $a=2/\Lambda$ can be interpreted as the effective radius of our point inclusions. 

\textit{Fluctuations.}---Using the stress tensor provides a way to calculate the variance of $f_x$:
\begin{eqnarray}
\Delta f_x^2&=& r^2 \int_\mathcal{D}d\alpha \,d\theta \,\Big[V_{xx}(\rv,\rv') \cos\theta \cos\alpha \nonumber\\
&&\!\!+ 2 V_{xy}(\rv,\rv') \cos\theta \sin\alpha +V_{yy}(\rv,\rv') \sin\theta \sin\alpha \Big]\!,\quad
\label{varforce}
\end{eqnarray}
where $V_{ij}(\rv,\rv')=\left\langle\Sigma_{xi}(\rv)\,\Sigma_{xj}(\rv')\right\rangle-\left\langle\Sigma_{xi}(\rv)\right\rangle\left\langle\Sigma_{xj}(\rv')\right\rangle$, with $\rv= r\,(\cos\theta,  \sin\theta)$, $\rv'= r\,( \cos\alpha,  \sin\alpha)$, and $\mathcal{D}=[0,2\pi]\times[0,2\pi]$.

Given the partition function (\ref{Z}), we may use Wick's theorem to express the correlation functions $V_{ij}$ in terms of $C$. For instance, the term in $V_{xx}(\rv,\rv')$ that originates from the product of the first and second terms in $\Sigma_{xx}$ gives
$v_{xx}(\rv,\rv')=-\frac{1}{2}\kappa^2\left[\partial_y^2\partial_{x'}^2 C(\rv,\rv')\right]^2$. We represent diagrammatically its contribution to $\Delta f_x^2$ as
\begin{equation}
r^2\!\int_\mathcal{D}\!d\alpha \,d\theta \,v_{xx}(\rv,\rv') \cos\theta \cos\alpha =-\frac{\kappa^2}{2}\diag{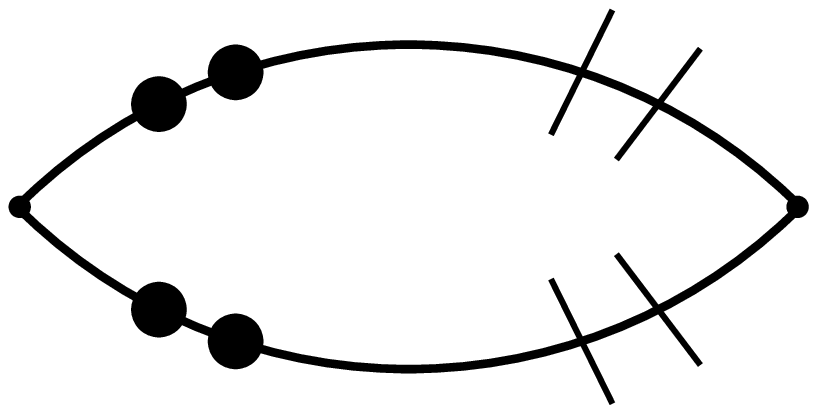}.
\label{diagr}
\end{equation}
Each line stands for the propagator $C(\rv,\rv')$, and the dashes (resp.\ dots) come from $\partial_x$ or $\partial_{x'}$ (resp.\ $\partial_y$ or $\partial_{y'}$) depending on whether they stand on the left or on the right of the diagram; the angular integrations are understood. All the diagrams that contribute to $\Delta f_x^2$ are displayed in the supplementary material \cite{SupMat}.
Since $C=G+G'$, each diagram breaks into four terms. For the diagram in Eq.~(\ref{diagr}), one of these terms reads explicitly:
\begin{eqnarray}
\diag{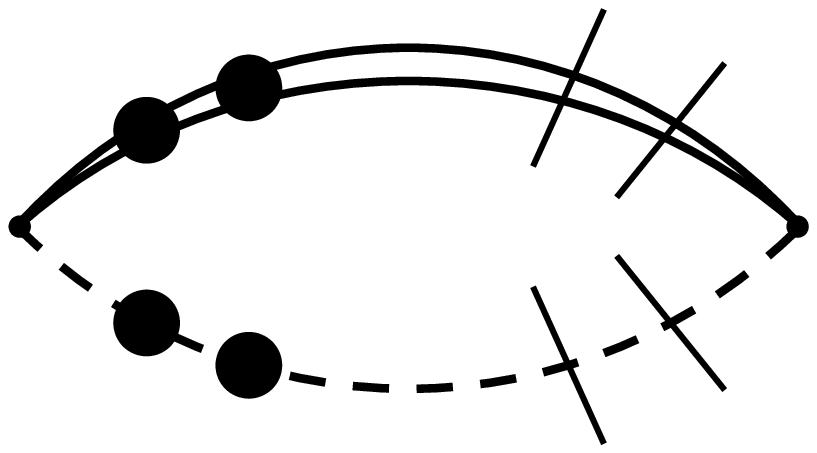}&=&- r^2 \!\int_{\qv,\kv,\qv'}\!\!\!\!\frac{q_y^2 k_x^2 \,\Delta_i(\qv)M^{-1}_{ij}\Delta_j(\kv)}{q^4 k^4}\,\frac{q'^2_y q'^2_x}{q'^4}\nonumber\\
&&\!\!\!\!\times\int_\mathcal{D}d\alpha \,d\theta \cos \theta \cos \alpha\, e^{i(\qv+\qv')\cdot\rv}e^{i(\kv-\qv')\cdot\rv'}\!,\qquad
\label{diag2}
\end{eqnarray}
where the double line stands for $G'(\rv,\rv')$ while the dashed line stands for $G(\rv-\rv')$, and $\Delta_j(\qv)=Q_j e^{-i \qv\cdot\Rv_j}$ with $Q=(q^2_x$, $q_x q_y$, $q^2_y$, $q^2_x$, $q_x q_y$, $q^2_y)$. 
The integrals in Eq.~(\ref{diag2}) can be performed analytically in the unphysical limit $\Lambda r\ll 1$. The result will guide our understanding of the physical case $r=a$. Summing all the diagrams in $\Delta f_x^2$ \cite{SupMat} then yields
$\Delta f_x^2\sim r^4\Lambda^6(k_\mathrm{B}T)^2\left[\ln\left(\Lambda\xi\right)-4(\Lambda d)^{-2}+\mathcal{O}\left((\Lambda d)^{-4}\right)\right]/192$, where zero-average oscillations at the cutoff frequency with amplitude in $(\Lambda d)^{-5/2}$ have been discarded.

\begin{figure}[tb]
  \begin{center}
    \includegraphics[width=\columnwidth]{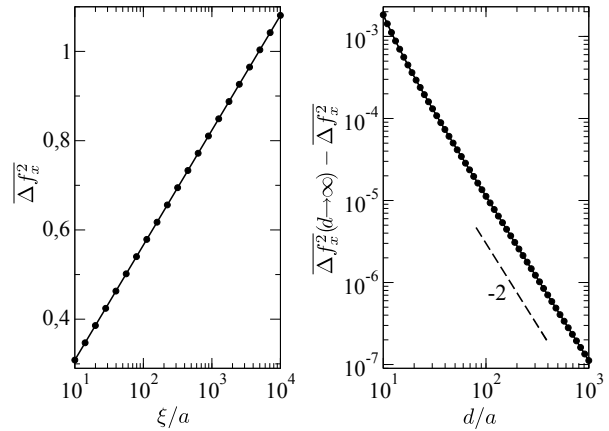}
    \caption{Left: Non-dimensional force variance $\overline{\Delta f_x^2}=\beta^2 a^2\Delta f_x^2$ for $d\to\infty$ as a function of $\xi/a$. The points are results from our numerical integrations, the line is the fit $\overline{\Delta f_x^2}=0.112 [\ln(2\xi/a)-0.239]$. Right: $d$ dependence of $\overline{\Delta f_x^2}$ for $\xi/a=10^3$. Oscillations at the cutoff frequency have been smoothed out. The line is the fit $\overline{\Delta f_x^2}(\!d\!\!\to\!\!\infty\!)-\overline{\Delta f_x^2}=0.112 \,a^2/d^2+7.18\, a^4/d^4$. Similar results were obtained for other values of $\xi/a$ ($10^2$ and $10^4$), with the same coefficient in front of $a^2/d^2$.}
  \label{Phi}
  \end{center}      
\end{figure}

Since even a piece of inclusion-free membrane is subject to a fluctuating force of finite variance, $\Delta f_x^2$ depends on the contour chosen to calculate $f_x$, contrary to $\left\langle f_x\right\rangle$. Thus, in order to obtain the fluctuation of the Casimir-like force acting on an inclusion, we should take a contour that includes the inclusion and only it. As in our model, the effective radius of the inclusions is $a=2/\Lambda$, the best we can do is to choose $r=a$ and to set, from now on, $\Lambda=2/a$. In order to calculate $\Delta f_x^2$ for $r=a$, we have computed the integrals of the diagrams numerically. Our study (see Fig.~\ref{Phi}) shows that the leading behavior of $\Delta f_x^2$ in $a/d$ is well described by the formula
\begin{equation}
\Delta f_x^2\sim0.112\frac{(k_\mathrm{B}T)^2}{a^2}\left[\ln\left(\frac{2\xi}{a}\right)-0.239-\frac{a^2}{d^2}\right],
\label{pa}
\end{equation}
which has the same main features as the analytical result found in the limit $\Lambda r\ll 1$. The only difference is the presence of the constant $-0.239$, which is negligible compared to the logarithmic term. As it can be absorbed in the definition of $\xi$, we shall discard it from now on. In the physical regime, corresponding to $\xi>>a$ and $d>2a$, (\ref{pa}) is always positive.

Our result (\ref{pa}) shows that $\Delta f_x^2$ is dominated by a distance-independent term, which corresponds to the fluctuations of the zero-average force exerted on a single inclusion by the membrane bulk. To leading order in $a/d$, the signal-to-noise ratio for the Casimir force is thus
\begin{equation}
\frac{\left\langle f_x \right\rangle}{\Delta f_x }\sim24\left[0.112\ln\left(\frac{2\xi}{a}\right)\right]^{-\frac{1}{2}} \left(\frac{a}{d}\right)^5,
\end{equation}
so that the Casimir force is very small compared to its fluctuations in the physical case $d>2 a$. Besides, the distance dependence of $\Delta f_x^2$ originates from the suppression of fluctuation modes by the boundary conditions imposed by the two inclusions, so it shares a common physical origin with the Casimir effect. To extract this $d$ dependence, we can define the Casimir effect \textit{relative to the fluctuations} as
\begin{equation}
\frac{\partial\Delta f_x^2}{\partial d}\sim0.224 \frac{(k_\mathrm{B}T)^2}{d^3}.
\end{equation}
This result, obtained for point inclusions, is independent of the cutoff. The question whether this universality holds for extended domains with dimension $a$ independent of the cutoff $\Lambda$ would be interesting to address.

\textit{Inclusions inducing a non-zero curvature.}---When the inclusions impose a finite local curvature (see, e.g., Fig.~\ref{BAR}), a component due to the average deformation of the membrane adds to the Casimir force \cite{Goulian93etc, Dommersnes99, Reynwar07}. The partition function can be obtained using the same method \cite{Dommersnes99, Dommersnes99b} as in the zero-curvature case discussed before. The correlation function $\langle h(\rv)\,h(\rv')\rangle-\langle h(\rv)\rangle\langle h(\rv')\rangle$ remains unchanged, equal to $C(\rv,\rv')$. The only change is the average deformation of the membrane: $\langle h(\rv)\rangle=V_i\,M^{-1}_{ij}\,\mathds{D}_j G(\rv-\Rv_j)$, where $V$ contains the elements of the curvature tensors of the two inclusions. We restrict ourselves to inclusions imposing the same isotropic curvature $c\ll a^{-1}$, so that $V=(c,0,c,c,0,c)$. The correspondence with finite-size inclusions imposing a contact angle variation $\alpha$ over a length $a$ is obtained for $c=\alpha/a$.

Integrating the average stress tensor over a contour that surrounds one inclusion but not the other (see Fig.~\ref{schema_casimir}) gives the sum of the Casimir force (\ref{FCas}) and a curvature-dependent force, i.e., 
\begin{equation}
\left\langle f_x\right\rangle= 24\,k_\mathrm{B}T \,\frac{(2/\Lambda)^4}{d^5}-32\pi\kappa\,\alpha^2  \,\frac{(2/\Lambda)^4}{d^5}+\mathcal{O}\left((\Lambda d)^{-7}\right).
\end{equation}
We thus recover from the stress tensor the total membrane-mediated interaction \cite{Goulian93etc, Park96, Dommersnes99}.

Studying the fluctuations of $f_x$ is a straightforward generalization of the zero-curvature case. The only difference is that Wick's theorem applies to $h(\rv)-\langle h(\rv) \rangle$ instead of $h(\rv)$, so that diagrams involving $\langle h(\rv) \rangle$ appear in the calculation. In the limit $\Lambda r\ll1$, we obtain analytically $\Delta f_x^2\sim r^4\Lambda^6(k_\mathrm{B}T)^2\,[\phi(\alpha)\ln(\Lambda\xi)-4\psi(\alpha)(\Lambda d)^{-2}+\mathcal{O}((\Lambda d)^{-4})]/192$ with $\phi(\alpha)=1+12\pi\beta\kappa\alpha^2$ and $\psi(\alpha)=1+8\pi\beta\kappa\alpha^2$.
The conclusions drawn in the vanishing curvature case still hold, but the Casimir effect relative to the fluctuations becomes
$\partial\Delta f_x^2/\partial d\sim\frac{1}{24}(\Lambda r)^4(k_\mathrm{B}T)^2\psi(\alpha)d^{-3}$. Thus, in the non-small curvature regime $\alpha>(8\pi\beta\kappa)^{-1/2}$, we obtain
\begin{equation}
\frac{\partial\Delta f_x^2}{\partial d}\propto\frac{k_\mathrm{B}T\kappa\alpha^2}{d^3}.
\end{equation}
The $k_\mathrm{B}T\kappa\alpha^2$ factor, which replaces the $(k_\mathrm{B}T)^2$ appearing in the pure Casimir case, reveals the interplay between the Casimir force and the curvature-dependent force.

In summary, the stress tensor is a powerful tool that allows for studying the Casimir-like force between membrane inclusions, especially its fluctuations. Using a coarse-grained description in which the inclusions are point-like, we have calculated the variance of the Casimir force. Our results show that the fluctuations dominate the average force, and that they depend on the distance between the inclusions. Further possible developments include treating the case of extended inclusions, and testing our results using coarse-grained membrane numerical simulations such as those in Ref.~\cite{Reynwar07}.

\end{document}